\documentclass[twocolumn,pra,aps,superscriptaddress,notitlepage]{revtex4-1}
\usepackage[utf8]{inputenc}

\usepackage{graphicx,xcolor}
\usepackage{amsmath,amssymb,amsthm,slashed,bm}
\usepackage{hyperref}

\def\<{\left<}
\def\>{\right>}
\def\ket|#1>{\left|#1\right>}
\def\bra<#1|{\left<#1\right|}
\def\elem<#1|#2|#3>{\left<#1\right|#2\left|#3\right>}
\def\[{\left[}
\def\]{\right]}
\def\bJ{\textbf{J}}
\def\bmu{\pmb{\mu}}
\def\nn{\nonumber}
\def\pl{\partial}

\def\Tr{\text{\rm Tr}}

\def\({\left(}
\def\){\right)}

\def\R{{\mathbb R}}

\def\ve{\varepsilon}
\def\tJ{\tilde{J}}
\def\tpsi{\tilde{\psi}}
\def\tPsi{\tilde{\Psi}}
\def\tx{{\tilde{x}}}

\def\beq{\begin{equation}}
\def\eeq{\end{equation}}

\begin{document}

\title[Short Title]{Depletion in fermionic chains with inhomogeneous hoppings}

\author{Begoña Mula}
\affiliation{Dto. Física Fundamental, Universidad Nacional de
  Educación a Distancia (UNED), Madrid, Spain}
\affiliation{Dto. Matemáticas, Universidad Carlos III de Madrid,
  Leganés, Spain}

\author{Nadir Samos Sáenz de Buruaga}
\affiliation{Instituto de Nanociencia y Materiales de Aragón (INMA), CSIC-Universidad de Zaragoza, Zaragoza, Spain} 

\author{Germán Sierra}
\affiliation{Instituto de Física Teórica UAM/CSIC, Universidad
  Autónoma de Madrid, Cantoblanco, Madrid, Spain}

\author{Silvia N. Santalla}
\affiliation{Dto. Física \&\ GISC, Universidad Carlos III de Madrid,
  Leganés, Spain}

\author{Javier Rodríguez-Laguna}
\affiliation{Dto. Física Fundamental, Universidad Nacional de
  Educación a Distancia (UNED), Madrid, Spain}

\date{November 23, 2022}

\begin{abstract}
  The ground state of a free-fermionic chain with inhomogeneous
  hoppings at half-filling can be mapped into the Dirac vacuum on a
  static curved space-time, which presents exactly homogeneous
  occupations due to particle-hole symmetry. Yet, far from
  half-filling we observe density modulations and depletion
  effects. The system can be described by a 1D Schrödinger equation on
  a different static space-time, with an effective potential which
  accounts for the depleted regions. We provide a semiclassical
  expression for the single-particle modes and the density profiles
  associated to different hopping patterns and filling
  fractions. Moreover, we show that the depletion effects can be
  compensated for all filling fractions by adding a chemical potential
  proportional to the hoppings. Interestingly, we can obtain exactly
  the same density profiles on a homogeneous chain if we introduce a
  chemical potential which is inverse to the hopping intensities, even
  though the ground state is different from the original one.
\end{abstract}

\maketitle

%%%%%%%%%%%%%%%%%%%%%%%%%%%%%%%%%%%%%%%%%%%%%%%%%%%%%%%%%%%%%%%%%%%

\section{Introduction}
\label{sec:intro}

Free fermionic chains are one of the most relevant basic models of
quantum many-body physics. Beyond its relevance in condensed matter
physics, they constitute one of the basic structures behind quantum
simulators \cite{Schafer.20,Gross.17}, which promise to help
understand many interesting phenomena. For example, fermionic chains
have been put forward to simulate the Dirac vacuum in curved
space-times, which would lead us to perform experiments on the Unruh
effect or Casimir forces on a background gravitational field
\cite{Boada.11,Laguna.17b,Bego.21}. Such quantum simulators can be
built using ultracold fermionic atoms on an optical lattice, employing
modulated laser beams to provide inhomogeneous hopping amplitudes
between neighboring cells \cite{Lewenstein.12}. The key insight is
that an inhomogeneity in the hoppings will give rise to an effective
space-time metric in the thermodynamic limit, under some mild
conditions.

\begin{figure}
  \includegraphics[width=7.8cm]{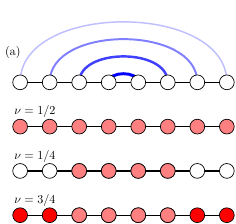}
 \vskip 5mm
  \includegraphics[width=7.8cm]{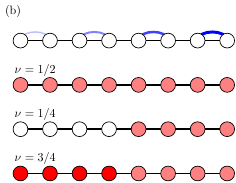}
  \caption{Depletion in free-fermionic chains with inhomogeneous
    hoppings, explained using the SDRG for different values of the
    filling fraction $\nu$. In (a) we have a rainbow chain whose
    single-particle orbitals are bonds between symmetrically placed
    sites. For $\nu=1/2$, all the bonds get single occupation (light red) and the
    density is exactly homogeneous. For $\nu=1/4$ we only occupy the
    two strongest bonds, leaving a depleted area near the borders. For
    $\nu=3/4$ the weakest bonds get double occupation (bright red), while the
    strongest remain with single occupation. In (b) we have a
    dimerized chain such that the energy associated to each bond grows
    rightwards. For $\nu=1/2$ we obtain the same homogeneous
    density. For $\nu=1/4$ only the rightmost bonds are occupied, and
    the left half is depleted. For $\nu=3/4$ the leftmost bonds obtain
    double occupation, and the rightmost ones still get one particle.}
  \label{fig:illust}
\end{figure}

In the aforementioned examples the Dirac vacuum is obtained as the
ground state (GS) of the lattice Hamiltonian at half-filling.
Interestingly, when the underlying lattice is bipartite the system
presents particle-hole symmetry and the occupation numbers become
exactly homogeneous. Moreover, its large scale physical properties can
be accounted for using conformal invariance arguments on the
appropriately deformed metric
\cite{DiFrancesco,Mussardo,Laguna.17,Tonni.18,MacCormack.18}. Yet, as
the filling fraction is lowered (or raised) the density will vary from
point to point. Morever, it may become negligible in the region
containing the lowest hopping amplitudes, a phenomenon which we have
termed {\em depletion}. This result can be readily understood in the
strong inhomogeneity regime, employing a strong-disorder
renormalization group (SDRG) scheme \cite{Dasgupta.80}, because the
orbitals with the lowest energies are localized upon the lowest
hopping amplitudes, which may correspond sometimes to effective
long-distance renormalized bonds. An illustration of this situation
can be seen in Fig. \ref{fig:illust}. Yet, in the weak inhomogeneity
limit the mathematical description of the depletion effects faces some
technical challenges: second-order derivatives of the fields must be
considered in the gradient expansion of the Hamiltonian, thus breaking
explicitly the conformal symmetry which characterizes the half-filling
case. 

This depletion has already been observed by other authors. For
example, it has been reported that the entanglement entropies of
inhomogeneous fermionic chains away from half-filling can be
interpreted as those corresponding to an effective shorter chain,
corresponding to the non-depleted region \cite{Finkel.21}. Moreover,
the effect of a finite density fermion field in entanglement has been
studied both in the relativistic \cite{Daguerre.21} and
non-relativistic frameworks \cite{Mintchev.22}.

This work addresses the emergence of depletion in inhomogeneous
pure-hopping free fermionic chains away from half-filling, and it is
divided as follows. In Sec. \ref{sec:model} we describe our physical
model, while Sec. \ref{sec:sdrg} describes the depletion phenomenon in
the strong disorder limit. Sec. \ref{sec:climit} describes our
continuum approximation for all filling fractions, and the effective
Schrödinger equation on a different space-time metric.
Sec. \ref{sec:density} describes our theoretical approach to the
density profiles and the depleted areas. The question of the effective
potential is addressed in Sec. \ref{sec:potential}, showing that it
can be either proportional to the hopping amplitudes or inversely
proportional to them, depending on our precise definition. Finally,
Sec. \ref{sec:conclusions} summarizes our findings and discusses our
suggestions for further work.

%%%%%%%%%%%%%%%%%%%%%%%%%%%%%%%%%%%%%%%%%%%%%%%%%%%%%%%%%%%%%%%%%%%%

\section{Model}
\label{sec:model}

Let us consider an open fermionic chain with $N$ (even) sites, whose
Hilbert space is spanned by creation operators $c^\dagger_i$, $i\in
\{1,\cdots,N\}$ following standard anticommutation relations,
$\{c^\dagger_i,c_j\}=\delta_{i,j}$. We define an inhomogeneous
hopping Hamiltonian,

\beq
H(\bJ)_N=-\sum_{i=1}^{N-1} J_i\; c^\dagger_i c_{i+1} + \text{h.c.},
\label{eq:ham}
\eeq
where $\bJ=\{J_i\}_{i=1}^{N-1}$ are the {\em hopping amplitudes},
$J_i\in \R^+$ referring to the link between sites $i$ and $i+1$. The
eigenstates of \eqref{eq:ham} can be obtained easily diagonalizing the
hopping matrix $J$, $J_{ij}\equiv J_i \(\delta_{i,j+1}+\delta_{i,j-1}\)$. So we can write $J=U\ve U^\dagger$,
where $\ve$ is a diagonal matrix whose $i$-th entry is $\ve_i$, the
single-body energy associated to the orbital given by the $i$-th
column of matrix $U$. We perform a canonical transformation,
$b^\dagger_k = \sum_i U_{ki} c^\dagger_i$, such that

\beq
H(\bJ)=\sum_{k=1}^N \ve_k b^\dagger_k b_k,
\eeq
where $\ve_k$ are arranged in ascending order, and we can write a basis of eigenstates of $H(\bJ)$ by fixing the
occupation numbers of the $b^\dagger_k$ modes. Let us consider the
minimum energy eigenstate with a fixed number of particles $m$,
which is obtained by filling up the lowest $m$ single-particle
modes,

\beq
\ket|\psi_m>=\prod_{k=1}^m b^\dagger_k \ket|0>,
\eeq
where $\ket|0>$ is the Fock vacuum. The filling fraction is defined as
$\nu\equiv m/N$.

\bigskip

In all cases, we will assume that the sequence of hopping amplitudes
presents a proper thermodynamic limit. Let $\bJ_N=\{J_{i,N}\}_{i=1}^{N-1}$
be a family of hopping amplitudes for all possible chain lengths
$N$. Then, we assume that there exists a continuous function $J: [0,1]
\mapsto \R^+$ such that $J_{i,N}=J(i/N)$. For concreteness, let us
consider three different examples. The {\em Rindler metric} is the
spacetime structure perceived by an observer moving with constant
acceleration in a Minkowski metric, described by

\begin{equation}
  J(x)=J_0x.
  \label{eq:rindler_J}
\end{equation}
Another natural choice is the {\em sine metric},

\begin{equation}
  J(x)=J_0 + J_1\cos\(2\pi x\),
  \label{eq:sine_J}
\end{equation}
or the {\em rainbow metric}
\cite{Vitagliano.10,Ramirez.14,Ramirez.15,Laguna.16,Laguna.17,Tonni.18,MacCormack.18,Samos.19,Samos.20,Samos.21},
given by

\begin{equation}
  J(x)=J_0\exp\(-h\left|x-\frac{1}{2}\right|\),
  \label{eq:rainbow_J}
\end{equation}
for $h\geq 0$, with $h=0$ corresponding to the Minkowski case. This
metric presents a constant negative curvature except at the center
\cite{Ramirez.15,Laguna.17}, $x=1/2$, thus resembling an anti-de
Sitter (adS) space \cite{MacCormack.18}, and has been extensively
discussed because it presents a maximal apparent violation of the area
law of the entanglement entropy. Notice that the $J_0$ parameter is
irrelevant in all cases, since it just fixes the global energy scale,
and we will take it as one.

\subsection{Density and particle-hole symmetry}
\label{subsec:homog}

The correlation matrix can be easily computed for the GS of
Hamiltonian \eqref{eq:ham},

\beq
C_{ij}\equiv\bra<\psi_m|c^\dagger_i c_j\ket|\psi_m>=\sum_{k=1}^m \bar U_{ki} U_{kj},
\eeq
and, concretely, the local occupation or density is found as
$\<n_i\>=C_{ii}$. Since our system is bipartite, let us define an operator
$P$ acting on the single-particle wavefunctions that
flips the sign of all components within one of the sublattices. It is
easy to prove that $JP=-PJ$. In other terms, if $U_k$ is an eigenstate
of $J$ with energy $\ve_k$, then $PU_k$ will be another eigenstate of
$J$ with energy $-\ve_k$. Every negative energy orbital has a positive
energy partner related through a $P$ operation, thus proving the 
particle-hole symmetry of the spectrum. Since $U$ is a unitary matrix, 
$\sum_{k=1}^N |U_{ki}|^2=1$, for all $i$. We may decompose the sum into two,

\beq
1=\sum_{k=1}^{N/2} |U_{ki}|^2 + \sum_{k=N/2+1}^N |U_{ki}|^2,
\eeq
but the second sum must be exactly the same as the first, because
$|U_{k,i}|^2=|U_{N+1-k,i}|^2$. Therefore, each sum must add up to 1/2,
thus proving that the eigenstate $\ket|\psi_{N/2}>$ at half-filling
must have homogeneous occupation, $n_i=1/2$ for all sites.

%%%%%%%%%%%%%%%%%%%%%%%%%%%%%%%%%%%%%%%%%%%%%%%%%%%%%%%%%%%%%%%%%%%%%%%

\section{Depletion at strong inhomogeneity}
\label{sec:sdrg}

\begin{figure}
  \includegraphics[width=8cm]{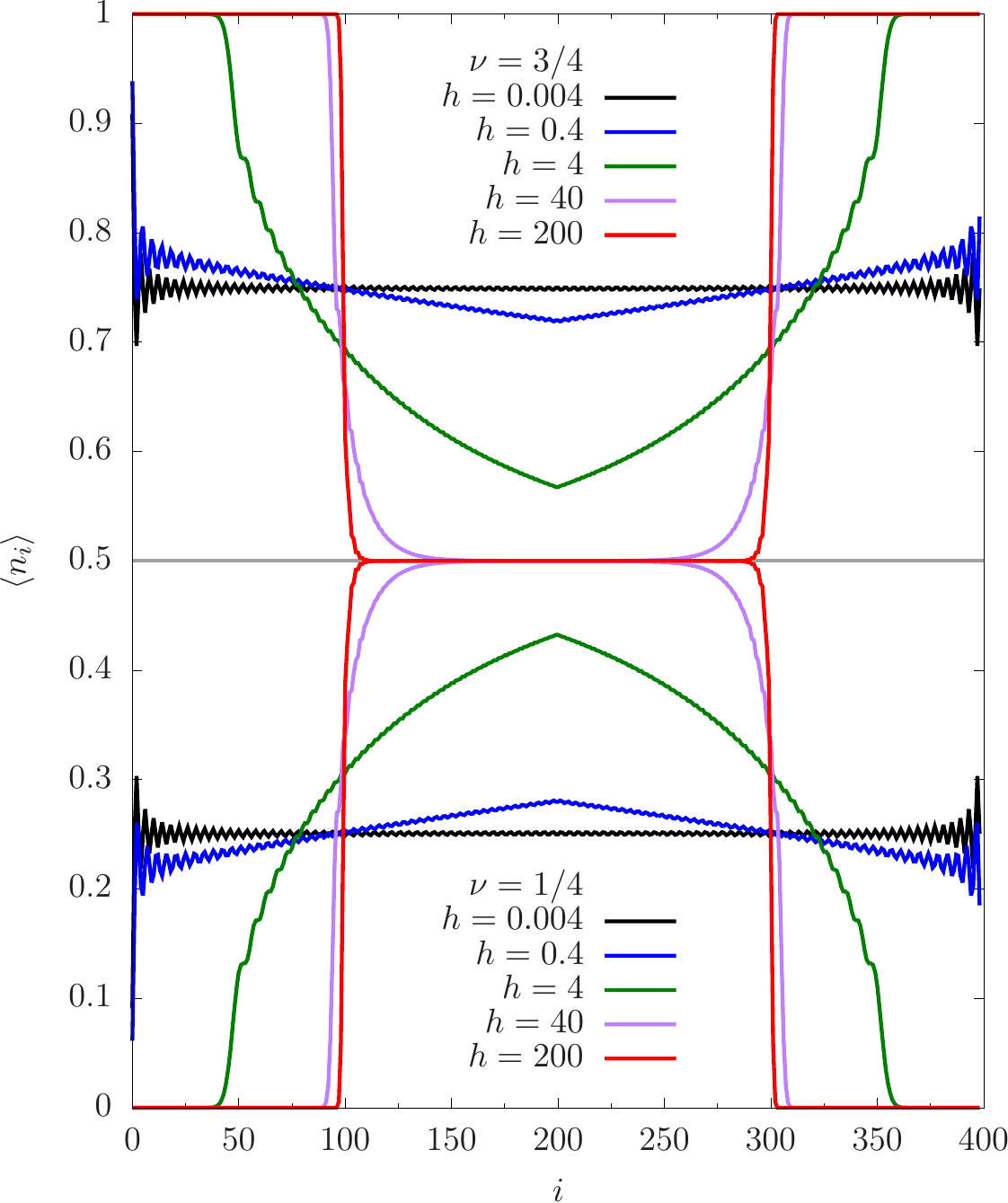}
%   \hspace{1cm}
%   \includegraphics[width=8cm]{fig_ni_Rb_varios34.pdf}
  \caption{Density profiles for the rainbow chain with $N=400$ sites
    and two filling fractions, $\nu=1/4$ and $\nu=3/4$, for different
    values of the inhomogeneity parameter $h$. Notice that for low
    inhomogeneities we always have a nearly flat profile, which gets
    more modulated as the inhomogeneity increases, converging towards
    the SDRG prediction as the inhomogeneity becomes large. As predicted by particle-hole
    symmetry,
    the $\nu=1/4$ and $\nu=3/4$ cases present mirror symmetry for all
    the values of the inhomogeneity.}
  \label{fig:strong}
\end{figure}

Let us consider the strong inhomogeneity regime, in which the values
of the hopping amplitudes differ largely between different links. In
this situation, the strong-disorder renormalization group (SDRG)
approach developed by Dasgupa and Ma describes the low-energy states
very effectively
\cite{Dasgupta.80,Refael.04,Laflorencie.05,Hoyos.07,Ramirez.14b}. Indeed,
the SDRG algorithm instructs us to select the most energetic link,
$J_i$, and to establish a bonding or anti-bonding orbital over the
corresponding couple of sites, depending on the hopping sign, which
are afterwards removed from the system.  Let us stress that the
particle occupying such a bond has the same probability of being found
on each site. The neighboring sites to this new bond are then linked
among themselves by an effective hopping amplitude, which is obtained
via second-order perturbation theory,

\beq
\tilde J_i = -\frac{J_L J_R}{J_i},
\eeq
where $J_L$ and $J_R$ are the left and right neighboring hopping
amplitudes, and the minus sign is due to the fermionic nature of the
particles. Notice that this new effective link can be selected in the
next iteration, if it happens to be the strongest one, thus yielding a
long-distance bond. Some interesting synthetic states, such as the
rainbow state, are built in such a way that all bonds (except the
first one) are long-distance \cite{Vitagliano.10,Ramirez.14}.

When the SDRG algorithm is performed at half-filling we fill up
$m=N/2$ bonds, each of which delocalizes a particle between a
different pair of (perhaps not neighboring) sites. Thus, each site has
an occupation probability of $1/2$, in accordance with the theorem of
Sec. \ref{subsec:homog}. Yet, if we place $m<N/2$ particles, they will
always occupy the region with highest hopping amplitudes, even if the
renormalization procedure yields long-distance
bonds. Fig. \ref{fig:strong} shows indeed that, in the strong
inhomogeneity regime, we can observe two different density regions for
$m<N/2$: half-occupied and empty. For $m>N/2$, due to particle-hole
symmetry, we occupy the same bonds but in reversed order. Each site
within a doubly occupied bond gets maximal occupation,
$\<n_i\>=1$. Therefore, as we increase the filling fraction above
$\nu=1/2$ we fill up completely the regions with the lowest hopping
amplitudes, mirroring the previous process.

This filling sequence is illustrated in Fig. \ref{fig:illust} (a) for
a rainbow chain and (b) a dimerized chain for filling
fractions $\nu=1/2$, $\nu<1/2$ and $\nu>1/2$ with a small size. In
Fig. \ref{fig:strong}, on the other hand, we can see the actual
densities numerically obtained for a larger chain, with $N=400$, using
$\nu=1/4$ and $\nu=3/4$, as a function of the parameter
$h$ which controls the level of inhomogeneity. As $h$ grows, we move
from a nearly uniform density profile towards the square profile of
the SDRG prediction, which differs in the cases of $\nu=1/4$ and
$\nu=3/4$. Indeed, in both cases we have $\<n_i\>=1/2$ at the central
half of the chain, while the lateral regions are depleted for
$\nu=1/4$ or full for $\nu=3/4$.

%%%%%%%%%%%%%%%%%%%%%%%%%%%%%%%%%%%%%%%%%%%%%%%%%%%%%%%%%%%%%%%%

\section{Depletion at weak inhomogeneity}
\label{sec:climit}

In this section we establish a continuum approximation to Hamiltonian
\eqref{eq:ham} for all possible filling fractions obtained through a
gradient expansion. At half-filling our model is known to map into the
Dirac Hamiltonian in the continuum limit,

\beq
i\slashed{D}_x \psi(x) =0,
\label{eq:masless_dirac}
\eeq
where $D_x$ is the covariant derivative on a given space-time metric
given by the hopping function $J(x)$ \cite{Ramirez.15,Laguna.17}. In
this work we will consider the situation away from half-filling,
showing that the gradient expansion should be taken to second order in
the derivatives of the field, giving rise to a continuum approximation
based on the Schrödinger equation on a curved background metric,

\beq
-\nabla^2_x \psi(x) + V(x)\psi(x)= E\psi(x),
\label{eq:schr}
\eeq
where $\nabla_x^2$ stands for the {\em Laplace-Beltrami} operator on a
different manifold \cite{Rosenberg}, whose metric is also given by the
hopping function $J(x)$.

\medskip

\subsection{Dirac Hamiltonian}

Let us assume that the local creation and annihilation operators can
be approximated in terms of two slowly varying fermionic fields,
$\psi_L(x)$ and $\psi_R(x)$, which makes reference to the right and left parts
of the wavefunction,

\begin{align}
c_m   &= \sqrt{a}\(e^{i k_F x} \psi_L (x)+e^{-i k_F x} \psi_R (x) \),\nn\\
c_m^\dagger &= \sqrt{a}\(e^{-i k_F x} \psi_L^{\dagger} (x)+
e^{i k_F x} \psi_R^\dagger (x) \),
\end{align}
where $a$ is the lattice spacing and $k_F$ the Fermi momentum, giving rise to the Hamiltonian

\begin{widetext}

\begin{equation}
\begin{split}
    \label{eq:ham_sin_expandir}
    H(x)=-\int_{0}^{\mathcal{N}} dx\; J(x)\[e^{-i k_F a} \psi^{\dagger}_L (x+a) \psi_L(x)+e^{i k_F a}\psi_R^{\dagger}(x+a)\psi_R(x) \right.\\
    \left. - e^{-i k_F a} e^{-2ik_Fx}\psi_L^{\dagger}(x+a)\psi_R(x)+e^{i k_F a} e^{2ik_Fx}\psi_R^{\dagger}(x+a)\psi_L(x)\],
\end{split}
\end{equation}
with $\mathcal{N}=Na$.

We should remark that the crossed terms, such as
$\psi_L^\dagger(x+a)\psi_R(x)$, have strongly oscillating prefactors
$e^{-2ik_F x}$, and thus their integral becomes negligible. As an
initial approach, we may expand the fields to first order in $a$,
$\psi(x+a)\approx \psi(x)+a\pl_x \psi(x)$, thus yielding the effective
Hamiltonian

%\beq
%\text{Final first-order Hamiltonian in derivatives}
%\eeq
%

\begin{equation}
\begin{split}
    \label{eq:ham_a}
  H(x)=-\int_{0}^{\mathcal{N}} dx J(x)\[ \( 2\cos(k_F a)- a e^{-i k_F a} \dfrac{J'(x)}{J(x)} \) \psi^{\dagger}_{L}(x) \psi_L(x) \right.\\
\left. + \( 2\cos(k_F a)- a e^{i k_F a} \dfrac{J'(x)}{J(x)} \) \psi^{\dagger}_{R}(x) \psi_R(x) \right.\\
\left. + 2 a i \sin(k_F a) \( \psi^{\dagger}_{L}(x) \partial_x \psi_L(x) - \psi^{\dagger}_{R}(x) \partial_x \psi_R(x) \) \].
\end{split}
\end{equation}
Let us perform a generic coordinate transformation, $x\to \tilde x$, such that 

\begin{equation}
\label{eq:transformation}
{d\tx\over dx} = \tilde{G}(\tx),
\end{equation}
so that $\pl_x =\tilde{G}(\tx)\pl_{\tx}$, $\pl_x^2
=\tilde{G}(\tx)\tilde{G}'(\tx) \pl_{\tx}+\tilde{G}^2(\tx)\pl_{\tx}^2$
and $\psi(x)=\tpsi(\tx)\tilde{G}^{1/2}(\tx)$. We will choose $\tilde{G}(\tx)$ so as to make the coefficient of the first derivative homogeneous. Therefore, $\tilde{G}(\tx)=\tilde{J}(\tilde{x})$ and Eq. \ref{eq:ham_a} can be written as

\begin{equation}
\begin{split}
    \label{eq:ham_a_tilde}
    H(\tilde{x})=-\int_0^{\mathcal{N}} d\tilde{x} \[2ai\sin(k_Fa)\( \tilde{\psi}_L(\tilde{x})^{\dagger} \partial_{\tilde{x}} \tilde{\psi}_L(\tilde{x})-\tilde{\psi}_R^{\dagger}(\tilde{x}) \partial_{\tilde{x}} \tilde{\psi}_R(\tilde{x}) \) \right.\\
    \left. + \cos(k_Fa) \(2 \tilde{J}(\tilde{x})-a\dfrac{\tilde{J}'(\tilde{x})}{\tilde{J}(\tilde{x})} \) \(\tilde{\psi}_L^{\dagger}(\tilde{x})\tilde{\psi}_L(\tilde{x})+\tilde{\psi}_R^{\dagger}(\tilde{x})\tilde{\psi}_R(\tilde{x})\) \],
\end{split}
\end{equation}
which in the case of half-filling, i.e. $k_Fa\to \pi/2$, reduces to

%\beq
%\text{Dirac Hamiltonian at half-filling.}
%\eeq
%
\begin{equation}
\label{eq:ham_Dirac}
    H_D(\tilde{x})=-\int_{0}^{\mathcal{N}} d\tilde{x}\;  2 a i\;  \( \tilde{\psi}^{\dagger}_{L}(\tilde{x}) \partial_{\tilde{x}} \tilde{\psi}_L(\tilde{x}) - \tilde{\psi}^{\dagger}_{R}(\tilde{x}) \partial_{\tilde{x}} \psi_R(\tilde{x}) \).
\end{equation}
Yet, we observe that for $k_Fa <\pi/2$ the Dirac equation acquires a
potential term, which may seem at first sight to be responsible for
the depletion effect, but is not. Indeed, the eigenstates of
\eqref{eq:ham_a_tilde} can be obtained in a closed form (equivalently
for $\tilde\psi_R$) by just solving the respective equations of motion,

%\beq
%\text{Solution of the Dirac equation,}
%\eeq

\begin{equation}
    \label{eq:dirac_sol}
    \tilde{\psi}_L(\tilde{x})=\exp{\[\dfrac{-i}{2a\sin(k_Fa)}\( \omega \tilde{x}-\int \cos(k_Fa) \(2\tilde{J}(\tilde{x})-a\dfrac{\tilde{J}'(\tilde{x})}{\tilde{J}(\tilde{x})}\) d\tilde{x} \) \]},
\end{equation}
where $\omega$ is an integration constant. In other words, the
wavefunctions are modulated plane waves in $\tx$ and they will not decay
exponentially.\\

\subsection{Second order approximation}
In order to reproduce the observed depletion effects we
should expand the fields to second order in the lattice parameter $a$,

\begin{equation}
  \psi(x+a)\approx \psi(x)+a \pl_x \psi(x)+ \dfrac{a^2}{2}\pl_x^2 \psi(x),
\end{equation}
thus yielding a Hamiltonian of the form

%\beq
%\text{Full Hamiltonian}
%\eeq
%

\begin{equation}
\begin{split}
    \label{eq:ham_full}
H(x)=-\int_{0}^{\mathcal{N}} dx J(x) \[ \( 2\cos(k_F a)- a e^{-i k_F a} \dfrac{J'(x)}{J(x)}+ \dfrac{a^2}{2} e^{-i k_F a} \dfrac{J''(x)}{J(x)} \) \psi^{\dagger}_{L}(x) \psi_L(x) \right.\\
+ \( 2\cos(k_F a)- a e^{i k_F a} \dfrac{J'(x)}{J(x)}+ \dfrac{a^2}{2} e^{i k_F a} \dfrac{J''(x)}{J(x)} \) \psi^{\dagger}_{R}(x) \psi_R(x)\\
+ 2 a i \sin(k_F a) \( \psi^{\dagger}_{L}(x) \partial_x \psi_L(x) - \psi^{\dagger}_{R}(x) \partial_x \psi_R(x) \) \\
+ a^2 e^{i k_F a} \dfrac{J'(x)}{J(x)} \psi^{\dagger}_{R}(x) \partial_x \psi_R(x) + a^2 e^{-i k_F a} \dfrac{J'(x)}{J(x)} \psi^{\dagger}_{L}(x) \partial_x \psi_L(x) \\
+ \left. a^2 \cos(k_F a) \( \psi^{\dagger}_{L}(x) \partial_{x}^{2} \psi_L(x)+\psi^{\dagger}_{R}(x) \partial_{x}^{2} \psi_R(x) \) \],
\end{split}
\end{equation}
which gives rise to the following equations of motion,

\begin{align}
\label{eq:eom}
E\psi(x)_{R/L}&=-J(x)\[\( 2\cos(k_Fa)-a e^{\pm ik_Fa}
  \dfrac{J'(x)}{J(x)}+\dfrac{a^2}{2} e^{\pm ik_Fa}
  \dfrac{J''(x)}{J(x)}\)\psi(x)_{R/L} \right. \nn\\ 
&  \left. \mp 2ai\sin(k_Fa) \pl_x \psi(x)_{R/L}+a^2 e^{\pm ik_Fa}
  \dfrac{J'(x)}{J(x)}\pl_x \psi(x)_{R/L} \right. \nn\\ 
&  \left. +a^2 \cos(k_Fa)\pl_x^2 \psi(x)_{R/L}  \],
\end{align}
where the $R/L$ notation makes reference to the right and left parts
of the wavefunction, and the corresponding sign should be chosen in
each case. We would like to notice that a second-order discrete
version of Eq. \eqref{eq:eom} with lattice spacing $a$ yields our
original single-body Hamiltonian \eqref{eq:ham}, thus proving the direct equivalence
between both systems.

\medskip

In the rest of the section we will transform this equation into a
Schrödinger equation, making use of two transformations: (a) a
coordinate transformation following Eq. \ref{eq:transformation}, which
is equivalent to embedding our system in a non-trivial space-time
metric, and (b) a gauge transformation in order to get rid of the
first derivative term.

\medskip

Our next purpose is then to make the coefficient of the second
derivative homogeneous through a suitable change of variable $\tx$.
Making a slight abuse of notation, we let $\tJ(\tx)\tilde{G}^2(\tx)=1$
$\rightarrow$ $\tilde{G}(\tx)=\tJ^{-1/2}(\tx)$. Notice the difference
with Eq. \ref{eq:ham_a}, in which we had $\tilde{G}(\tx)=\tJ(\tx)$
once the appropriate tranformation of coordinates was performed in
order to obtain a homogeneous first-derivative term. The equation of
motion in these new transformed coordinates reads

\begin{align}
\label{eq:e1}
E\tpsi(\tx)_{R/L} &
=-a^2 \cos(k_Fa)
\pl_\tx^2 \tpsi(\tx)_{R/L}
\pm 2i\sin(k_Fa)
\(a\tJ^{1/2}(\tx)-\dfrac{a^2}{2}
\dfrac{\tJ'(\tx)}{\tJ(\tx)}\)\pl_{\tx}
\tpsi(\tx)_{R/L}\nn\\
&-  2\cos(k_Fa) \( \tJ(\tx)-\dfrac{a}{4}
\dfrac{\tJ'(\tx)}{\tJ^{1/2}(\tx)}+\dfrac{7a^2}{32}
\dfrac{\tJ'^2(\tx)}{\tJ^2(\tx)}-
\dfrac{a^2}{8}\dfrac{\tJ''(\tx)}{\tJ(\tx)} \)
\tpsi(\tx)_{R/L}\nn\\
&-\dfrac{e^{\pm ik_Fa}}{2} \left(-a\dfrac{\tJ'(\tx)}{\tJ^{1/2}(\tx)} -a^2
\dfrac{\tJ'^2(\tx)}{\tJ^2(\tx)}+a^2\dfrac{\tJ''(\tx)}{\tJ(\tx)}
\right) \tpsi(\tx)_{R/L}.
\end{align}
This equation can be rewritten so as to make the single-body operator
manifestly hermitean,

\begin{align}
  E\tpsi(\tx)_{R/L} &
  =-a^2 \cos(k_Fa) \pl_\tx^2 \tpsi(\tx)_{R/L}\nn\\
& \pm 2i\sin(k_Fa) \(a \tJ^{1/4}(\tx)\pl_\tx\(\tJ^{1/4}(\tx)\tpsi(\tx)_{R/L}\)
  -\dfrac{a^2}{2}\(\dfrac{\tJ'(\tx)}{\tJ(\tx)}\)^{1/2}
  \pl_\tx\(\dfrac{\tJ'(\tx)}{\tJ(\tx)}\)^{1/2} \tpsi(\tx)_{R/L}\)\nn\\
& - 2\cos(k_Fa) \( \tJ(\tx)- \dfrac{a}{2}
  \dfrac{\tJ'(\tx)}{\tJ^{1/2}(\tx)}-\dfrac{a^2}{32}
  \dfrac{\tJ'^2(\tx)}{\tJ^2
    (\tx)}+\dfrac{a^2}{8}\dfrac{\tJ''(\tx)}{\tJ(\tx)}\)
  \tpsi(\tx)_{R/L}.
\end{align}
In order to transform our equation of motion into a Schrödinger equation, our next task is to get rid of the first derivative term
using a gauge-like transformation,

\begin{equation}
\label{eq:gaugetransf}
\tpsi(\tx)_{R/L}=e^{ i\beta_{R/L}(\tx)}\tPsi(\tx)_{R/L},
\end{equation}

\noindent which implies that

\begin{align}
  \pl_{\tx}\tpsi(\tx)_{R/L} &=
   i\beta_{R/L}'(\tx)e^{ i\beta_{R/L}({\tx)}}\tPsi(\tx)_{R/L}+e^{ i\beta_{R/L}(\tx)}\pl_{\tx}\tPsi(\tx)_{R/L},
  \nn\\ 
  \pl_{\tx}^2 \tpsi(\tx)_{R/L} &=
   i\beta_{R/L}''(\tx)e^{ i\beta_{R/L}(\tx)}\tPsi(\tx)_{R/L}-
  \beta_{R/L}'^2(\tx)e^{ i\beta_{R/L}(\tx)}\tPsi(\tx)_{R/L}
  2i\beta_{R/L}'(\tx)e^{
    i\beta_{R/L}(\tx)}\pl_{\tx}\tPsi(\tx)_{R/L}\nn\\
  &+
  e^{ i\beta_{R/L}(\tx)}\pl_{\tx}^2\tPsi(\tx)_{R/L}.
\end{align}
The condition that we have to impose so that the first-derivative terms cancel out is

\begin{equation}
  \beta_{R/L}'(\tx)=\mp \tan(k_Fa)
  \(\dfrac{1}{2}\dfrac{\tJ'(\tx)}{\tJ(\tx)}-\dfrac{1}{a}\tJ^{1/2}(\tx)\).
\end{equation}

\noindent Eq. (\ref{eq:gaugetransf}) is analogous to a gauge transformation where the coefficient of the first derivative term in Eq. (\ref{eq:e1}) plays the role of a gauge potential  $A(\tilde{x})$, up to a multiplicative factor. In this case, the change in the phase would be expressed by $\exp\({q\int_{0}^{\tilde{x}}A(u)du}\)$, where $q$ is a constant. Therefore,  $\beta '(\tilde{x})=A(\tilde{x})$.

\medskip

Applying this transformation we obtain 

\begin{align}
E\tPsi(\tx)_{R/L} &=- \dfrac{1+\cos^2(k_Fa)}{\cos(k_Fa)} \tJ(\tx)\tPsi(\tx)_{R/L}\nn\\
& + a \frac{1}{\cos(k_Fa)}\dfrac{\tJ'(\tx)}{\tJ^{1/2}(\tx)}
\tPsi(\tx)_{R/L} \nn\\
& -\dfrac{a^2}{4}
\(\dfrac{\sin^2(k_Fa)}{\cos(k_Fa)}
\dfrac{\tJ'^2(\tx)}{\tJ^2(\tx)}
-\dfrac{\cos(k_Fa)}{4}\dfrac{\tJ'^2(\tx)}{\tJ^2(\tx)}
+\cos(k_Fa)\dfrac{\tJ''(\tx)}{\tJ(\tx)}\)\tPsi(\tx)_{R/L}\nn\\
&-a^2 \cos(k_Fa) \pl^2_\tx \tPsi(\tx)_{R/L}.
\label{eq:schr_eff}
\end{align}

\end{widetext}

Eq. \eqref{eq:schr_eff} has the form of a Schrödinger equation in
$\tx$ with a mass $M={2/\cos(k_Fa)}$ that tends to zero as $k_Fa\to
\pi/2$, thus rendering the approximation invalid in that limit. The
effective potential, to zero order in $a$, becomes

\beq
V(\tx)\approx -\dfrac{1+\cos^2(k_Fa)}{\cos(k_Fa)} \tJ(\tx).
\label{eq:potential_zo}
\eeq
Notice that, for $k_Fa\ll 1$ we have to a very good approximation
$V(\tx)\approx -2 \tJ(\tx)$ or, equivalently, $V(x)=-2 J(x)$, while
for larger values of $k_Fa$ the different modes of our system
correspond to different Schrödinger equations. The reason is that the
prefactors of Eq. \eqref{eq:schr_eff} present explicit dependence on
$k_Fa$. The next corrections, corresponding to higher orders in $a$,
can be shown to be small or constant for the hopping amplitudes
employed in this work. We would also like to stress that our system is
now embedded on a manifold with metric $ds^2=dt^2-d\tx^2$, and all
geometrical measurements should be transformed back before further
comparisons with our original discrete model.

\bigskip

\begin{figure}
\includegraphics[width=7.5cm]{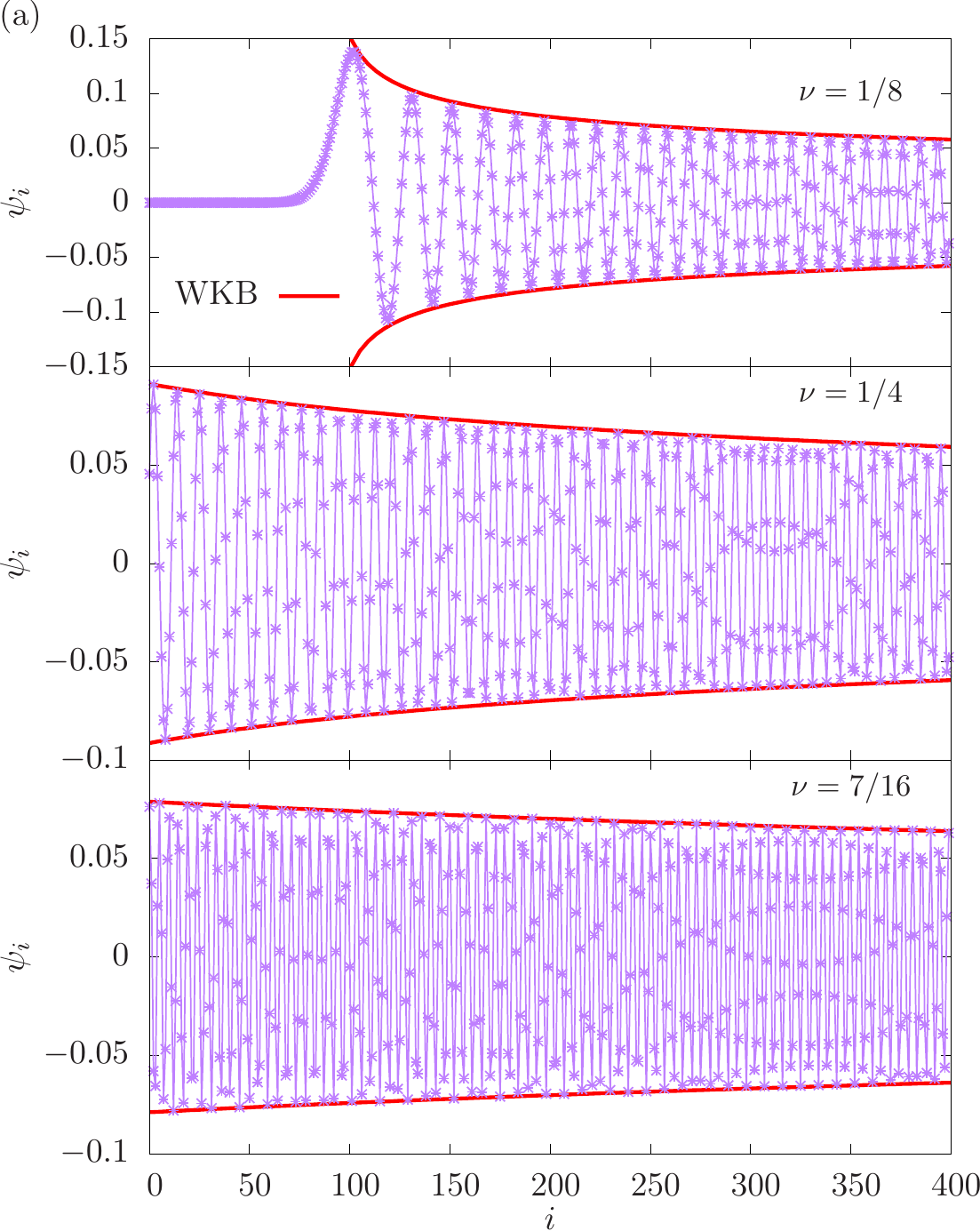}   
\includegraphics[width=7.5cm]{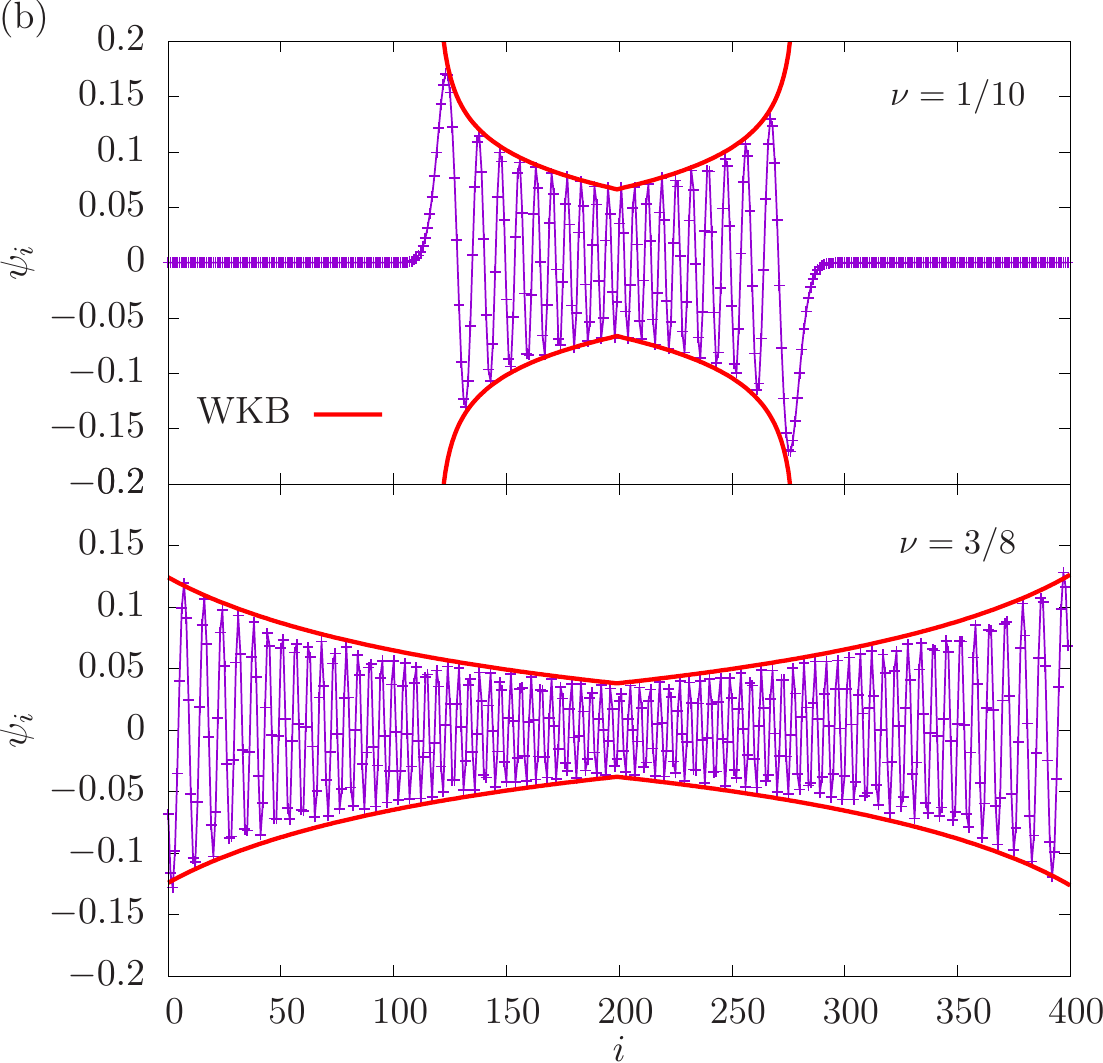}
   \caption{Comparing the modes obtained from Hamiltonian
    \eqref{eq:ham}, $\psi_i$, using $N=400$ for Rindler and
    Rainbow chains with the WKB approximation given in
    Eq. \eqref{eq:wkb}, considering different filling fractions $\nu$.
    The continuous red line represents the semiclassical
    approximation Eq. \eqref{eq:wkb} to the continuous approximation
    given in Eq. \eqref{eq:schr_eff}. Panel (a): Rindler chain, from top to bottom,
    $m=50$, 100 and 175. Panel (b): rainbow chain, top to bottom $m=40$
    and $m=150$.}
  \label{fig:modes}
\end{figure}

Let us discuss the numerical validity of Eq. \eqref{eq:schr_eff},
which is familiar to us due to its Schrödinger form. For highly
excited states we are allowed to perform a Wentzel-Kramers-Brillouin
(WKB) approximation, which leads to a form

\begin{equation}
    \label{eq:psi_oscilante}
    \tilde{\Psi}(\tilde{x})_{R/L}\sim
    \dfrac{1}{\sqrt{\tilde{p}(\tilde{x})}}
    e^{\pm i\tilde{p}(\tilde{x})\tilde{x}},
\end{equation}
in the new coordinate $\tx$, where $\tilde p(\tx)$ is the momentum of
a particle in that position according to classical mechanics,
i.e. $p(x)=\pm \sqrt{2M(E-V(x))}$

\begin{equation}
    \label{eq:p}
    \tilde p(\tx)=\sqrt{{1\over\cos^2(k_Fa)}\(\dfrac{1+\cos^2(k_Fa)}{\cos(k_Fa)} \tJ(\tx)+E\)},
\end{equation}
Yet, we should transform this solution back to our original coordinate
system in order to check its numerical validity, making use of the
change of coordinates for a probability distribution,
$|\tilde{\Psi}(\tilde{x})|^2 d\tilde{x}=|\Psi (x)|^2 dx$, which leads
to

\begin{equation}
  |\Psi(x)_{R/L}|=\dfrac{|\tilde{\Psi}(\tilde{x})_{R/L}|}{J^{1/4}(\tx)}
  \sim \dfrac{1}{\sqrt{\tilde{p}(\tilde{x})}} \dfrac{1}{J^{1/4}(\tilde{x})}.
    \label{eq:wkb}
\end{equation}
We have checked the validity of Eq. \eqref{eq:wkb} in
Fig. \ref{fig:modes}. In panel (a) we have chosen a Rindler system
with $J(x)=\frac{1}{4}+x$ with $N=400$ and shown the modes $m=50$, 100
and 175. The continuous red curve in each case corresponds to the
approximation \eqref{eq:wkb}, suitably normalized. We can see that the
decay is nearly perfect. Panel (b) shows the same situation for the
rainbow chain, using $h=4$, $N=400$, with the modes $m=40$ and
$m=150$. We can see that the decay is nearly perfect in all the cases.

We should stress that our continuum approximation, Eq. \eqref{eq:eom2}
can not be employed to obtain a continuum limit of our original
model. Indeed, for $a\to 0$, all terms containing derivatives of the
field $\psi(x)$ vanish, thus rendering the equation useless. In order
to make sense of Eq. \eqref{eq:eom2} we must keep $a$ finite and this
implies that we should preserve all derivatives of the gradient
expansion. Alternatively, we may define a new physical variable
$u=x/a$, in such a way that any derivative with respect to $x$
multiplied by $a$ becomes a derivative with respect to $u$:
$\pl_u=a\pl_x$.

%%%%%%%%%%%%%%%%%%%%%%%%%%%%%%%%%%%%%%%%%%%%%%%%%%%%%%%%%%%%%%%%%%%%%

\section{Density profiles}
\label{sec:density}

Our next aim will be to provide a continuum approximation to the
density profiles observed for free-fermionic chains with inhomogeneous
hopping amplitudes away from half-filling, based on the validity of
the Schrödinger equation on a different manifold,
Eq. \eqref{eq:schr_eff}, using a certain effective potential
$V(x)$. If we fill all orbitals up to a certain energy $E$, we we will
observe depletion in the classically forbidden regions, defined by
$V(x)>E$, and bounded by the turning points, defined by $V(x_*)=E$. To
order zero in $a$, from \eqref{eq:potential_zo}, we may estimate these turning points as

\beq
E=-\dfrac{1+\cos^2(k_Fa)}{\cos(k_Fa)} J(x_*).
\label{eq:turning_points}
\eeq
This result can also be obtained in a heuristic way, starting from a
simplified version of Eq. \eqref{eq:eom},

\begin{align}
E\psi(x)_{R/L}&=-J(x)\(2 \cos(k_Fa) \psi(x)_{R/L} \right. \nn\\
&\mp 2ai\sin(k_Fa)\pl_x \psi(x)_{R/L} \nn\\
&\left. +a^2 \cos(k_Fa) \pl_x^2 \psi(x)_{R/L}\).
\label{eq:eom2}
\end{align}
Now, we suppose that the wavefunction is locally a plane wave with a
certain position dependent momentum $q(x)$, i.e. $\psi(x)_{R/L} \sim
e^{\pm i q(x)x}$. Then,

\begin{align}
  E &=-J(x)(2 \cos(k_Fa) + 2a\sin(k_Fa)q(x)\nn\\
  &-a^2 \cos(k_Fa) q(x)^2),
\end{align}
and we can then obtain $q(x)$ solving a quadratic equation. If $q(x)$
is not real, then $x$ belongs to the classically forbidden
region. Thus, by making the discriminant zero we reach
Eq. \ref{eq:turning_points}.

\medskip

We can obtain an approximation to the local density $\rho(x)$ of a
Schrödinger equation by considering a particle with energy $E$
traveling through a small segment of size $\Delta x$ around position
$x$, where the potential energy is $V(x)$. Its momentum will be given
by

\beq
q(x)=\sqrt{2m(E-V(x))}.
\eeq
Within a semiclassical approximation we may estimate the number of
orbitals with presence on that segment assuming that the momenta are
discretized as $q(x)\approx n\pi/N\approx \pi\rho(x)$. Thus, we have

\beq
\rho(x)\approx {1\over \pi} \sqrt{2m\(E-V(x)\)}.
\eeq
In our case, for low values of $k_Fa$ we also build the density by
filling up the modes of a Schrödinger equation, and thus we may write

\beq
\tilde\rho(\tx) \approx {1\over\pi} \sqrt{{2\over
    a^2}\(E-2 \tJ(\tx)\)}.
\eeq
Yet, this expression is designed for the transformed coordinate
$\tx$. We should express it in our original coordinate in order to
make useful predictions, using $\tilde\rho(\tx)d\tx=\rho(x)dx$, we
have $\rho(x)=\tilde\rho(\tx) \tJ^{-1/2}(\tilde{x})$, and therefore

\beq
\rho(x) a \approx A \sqrt{ {E\over J(x)}-2},
\label{eq:rho}
\eeq
where $A$ is a normalization constant. Indeed, $\rho(x)a$ can be
interpreted as the local occupation, which can be directly compared to
$\langle c^\dagger_i c_i\rangle$ for $i=x/a$. Interestingly, the
density is directly related to the {\em inverse} of the hopping
function $J(x)$. Notice that Eq. \eqref{eq:rho} is not necessarily
valid for larger values of $k_Fa$, since the modes that we are filling
up do not correspond to the same Schrödinger equation. 

\bigskip

\begin{figure}
  \includegraphics[width=9cm]{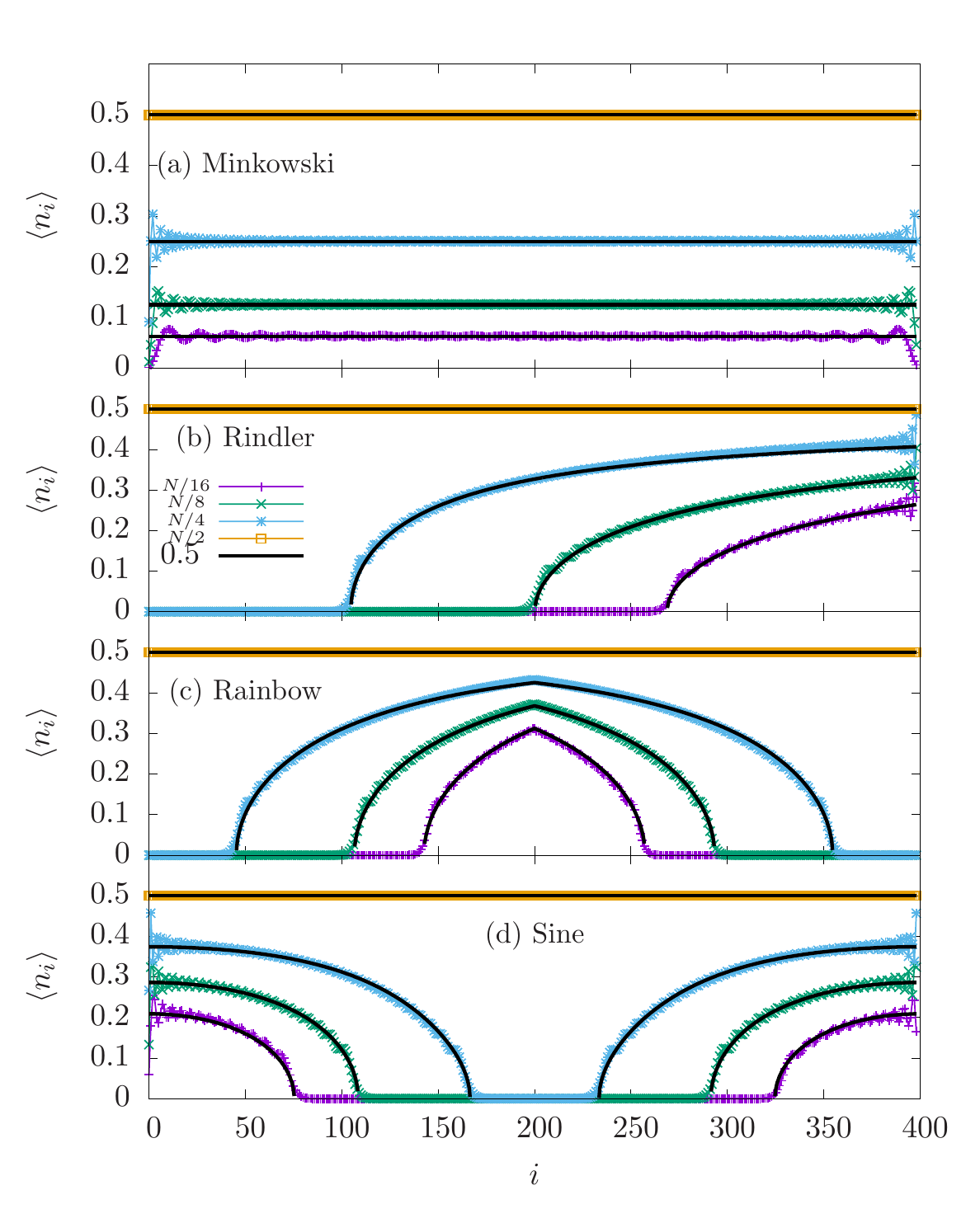}
  \caption{Fermionic density for a chain of $N=400$ sites with
    different filling fractions $1/2$, $1/4$, $1/8$ and $1/16$ and four
    different inhomogeneities: (a) Minkowski, (b) Rindler, $J(x)=x$,
    (c) Rainbow, $J(x)=\exp(-h|x-1/2|)$ with $h=0.01$ and (d) Sine, $J(x)=1+0.5\cos(2\pi x)$. The black curves correspond to the theoretical curves, given by Eq. \eqref{eq:rho}.
    }
  \label{fig:density_aFH}
\end{figure}

We have computed numerically the fermionic density for different
hopping functions, and observed depletion in all the considered cases,
except for the Minkowski spacetime, as we can see in
Fig. \ref{fig:density_aFH}. As expected, the depleted regions decrease
their size as the filling fraction grows. Moreover, Eq.
\eqref{eq:rho} predicts very well the density profiles for all $\nu
\leq 1/4$. Surprisingly, for all values of the filling fraction the
functional form

\beq
\rho(x)a = A \sqrt{{1\over J(x)}-B},
\eeq
fits extremely well the numerical density profiles, as we can check in
Fig. \ref{fig:density_aFH}.

%%%%%%%%%%%%%%%%%%%%%%%%%%%%%%%%%%%%%%%%%%%%%%%%%%%%%%%%%%%%%%%%%%%%%%%

\section{Compensating and mimicking potentials}
\label{sec:potential}

As we have discussed above, our continuum approximation led to an
effective Schrödinger equation with a potential whose classically
forbidden areas correspond to the depletion regions of the particle
density. 

\begin{figure}
  \includegraphics[width=8cm]{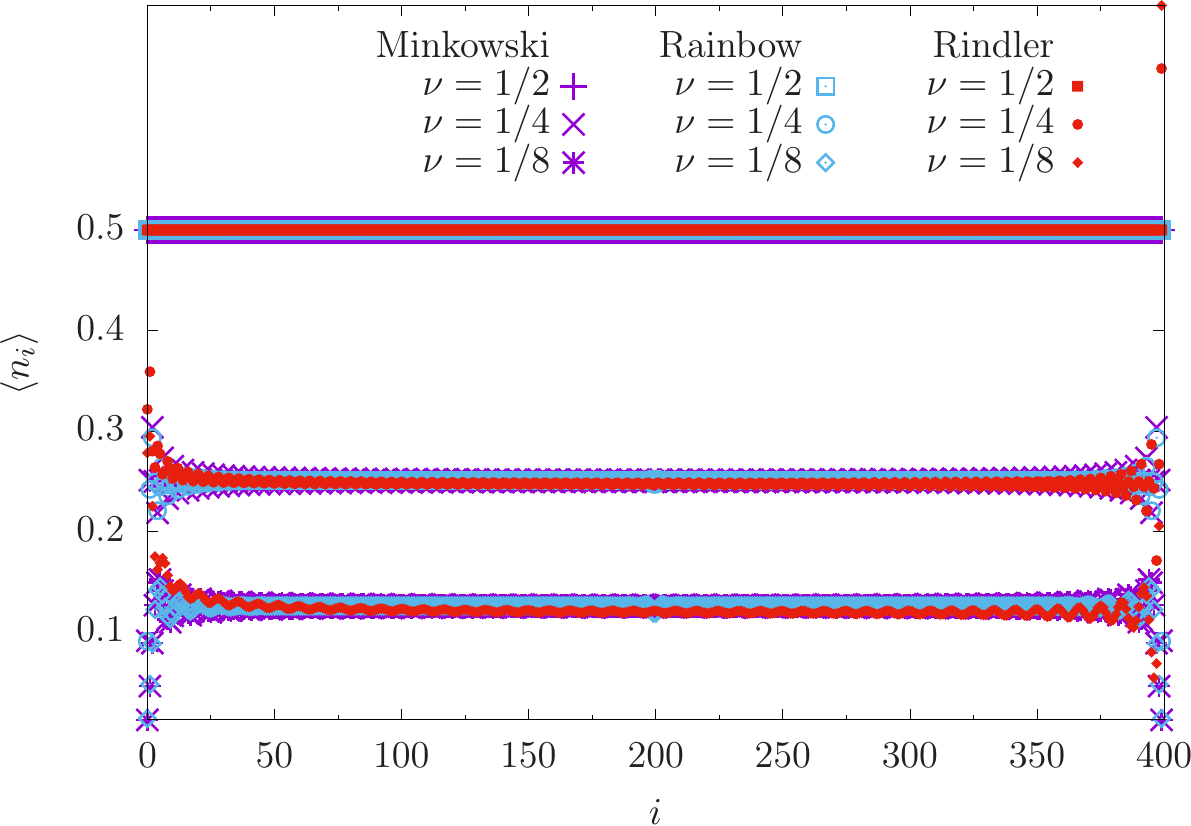}
  \caption{Density profile for the GS of Hamiltonian \eqref{eq:ham_mu} with a compensating potential,
    using three types of hopping functions: Minkowski, Rindler and
    rainbow, and different filling fractions, $\nu=1/2$, $1/4$ and
    $1/8$.}
  \label{fig:compensate}
\end{figure}

Let us extend our original model, Eq. \eqref{eq:ham}, introducing an
inhomogeneous chemical potential $\bmu=\{\mu_i\}_{i=1}^N$,

\beq
H(\bJ,\bmu)_N=-\sum_{i=1}^{N-1} J_i \(c^\dagger_i c_{i+1} +
  \text{h.c.}\)+  \sum_{i=1}^N \mu_i c^\dagger_i c_i.
\label{eq:ham_mu}
\eeq
We may introduce a {\em compensating potential}, defined by

\beq
\mu_i=\mu_0 J_i,
\eeq
where we implicitly assume that the chemical potential at site $i$ is
given by the average of its two neighboring hopping constants. With
such choice, the GS of Hamiltonian \eqref{eq:ham_mu} always presents a
flat density profile, for all filling fractions, whenever
$\mu_0=2\cos(k_Fa)$, as it can be checked in
Fig. \ref{fig:compensate}. In mathematical terms, the reason is that
the added chemical potential cancels out the potential energy term in
Eq. \eqref{eq:eom}. In this case the Hamiltonian does not present any
terms which are independent of the lattice spacing, $a$, and thus we
are allowed to renormalize the hopping function, $J(x)\to\infty$,
$a\to 0$, while $J(x)a\to \hat J(x)$, thus yielding a proper continuum
limit.

The physical meaning of this compensating effect is also
interesting. Let us start out with the GS of Hamiltonian
\eqref{eq:ham_mu} using $J_i=1$ and $\mu_i=0$, filled up with $\nu N$
fermions. We notice that the energy cost of introducing a new particle
does not decay to zero as the system size increases, and instead is
bounded by $-2\cos(k_Fa)$, with $k_Fa=\pi\nu$. Thus, the system can
not be conformally invariant. We can change that by introducing a
chemical potential, $\mu_i=2\cos(k_Fa)$. In that case, the energy cost
of introducing extra particles becomes zero. This new system can be
set in any different static 1+1D metric by introducing an appropriate
Weyl factor, thus yielding the compensating potential system
\cite{Laguna.17,Samos.21}.

\bigskip

Now we may ask a complementary question. Let us keep a flat hopping
function, $J(x)=1$, i.e. $J_i=1$. Can we find a chemical potential
$\{\mu_i\}$ which {\em mimics} the density profiles obtained from the
inhomogeneous hopping function without chemical potential, for the
same filling fraction? Interestingly, the answer is yes.

Let us consider the bulk equations to obtain the eigenstates of the
original hopping matrix. Let $(\psi_1\cdots\psi_N)^T$ be the
eigenvector of the hopping matrix with eigenvalue $E$. Then,

\beq
J_{n-1}\psi_{n-1}+J_n \psi_{n+1} = E\psi_n,
\eeq
which can be rewritten for very smooth $\bJ$ as

\beq
J_n (\psi_{n-1} + \psi_{n+1}) \approx E \psi_n.
\eeq
Now we can take the hopping amplitude to the RHS, assuming that it is
non-zero,

\beq
\psi_{n-1} + \psi_{n+1} -\({E\over J_n} -E\) \psi_n \approx E\psi_n,
\eeq
which can be read as a homogeneous hopping Hamiltonian with a chemical
potential $\mu_n$ of the form

\beq
\mu_n=E\({1\over J_n} - 1\),
\eeq
i.e. the chemical potential depends on the energy itself, and thus the
secular equation becomes non-linear. This reasoning motivates the
following {\em mimicking} chemical potential

\beq
\mu_i=\frac{\mu_0}{J_i}
\label{eq:inverse_mu}
\eeq

\begin{figure}
  \includegraphics[width=8cm]{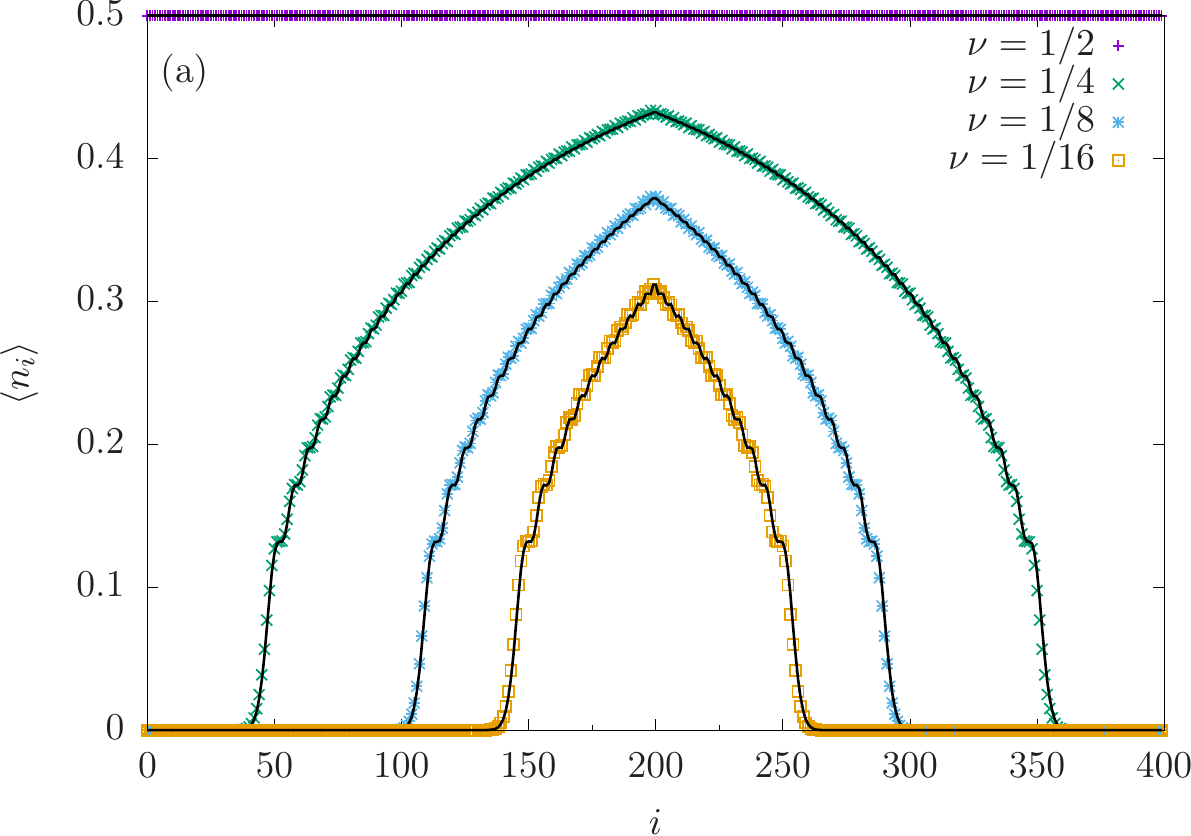}
  \includegraphics[width=8cm]{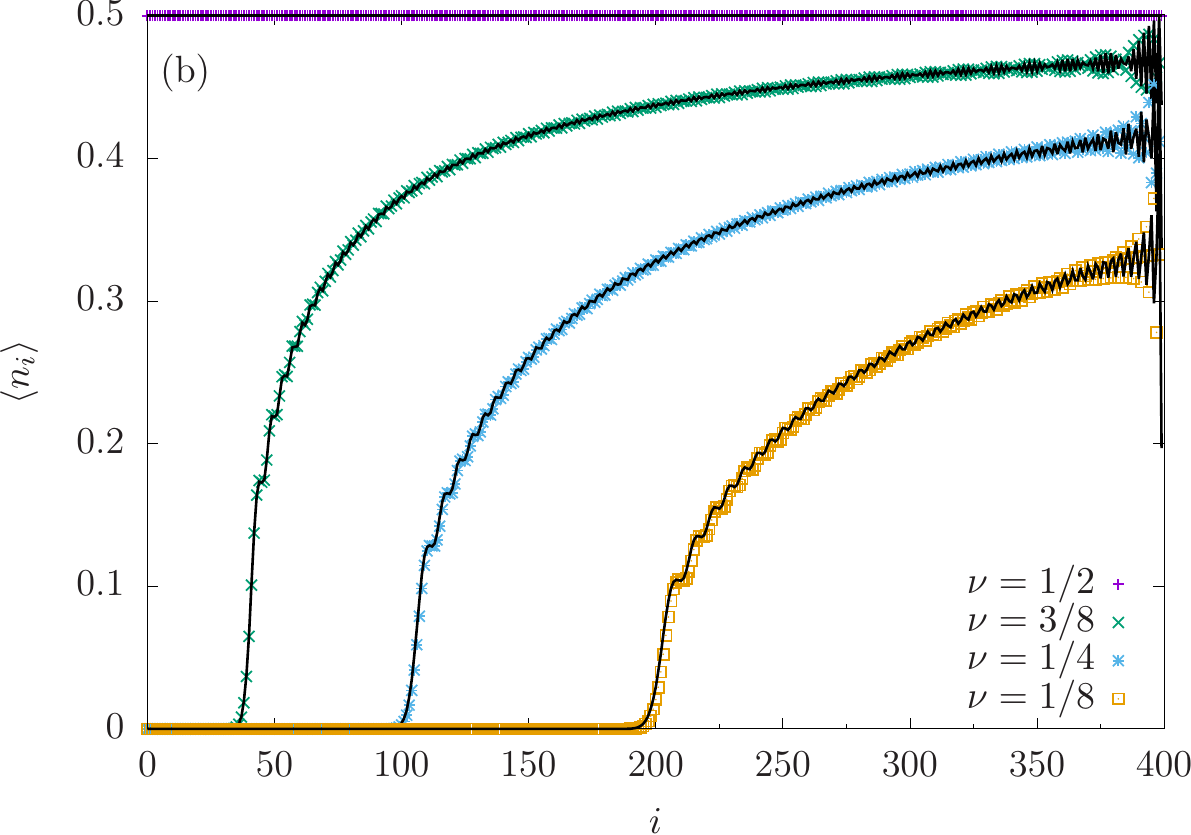}
  \caption{Density profiles of the GS of Hamiltonian \eqref{eq:ham_mu}
    using homogeneous hoppings, $J_i=1$, and the corresponding
    mimicking potential, Eq. \eqref{eq:inverse_mu}, for (a) a rainbow
    chain with $N=400$ and $h=4$ and (b) a Rindler system with
    $J(x)=x$, using the filling fractions shown in the key, along with
    the original density profiles using inhomogeneous hoppings and
    without chemical potential.}
    \label{fig:simulating}
\end{figure}

In Fig. \ref{fig:simulating} we plot the density profiles associated
to the ground states of Eq. \eqref{eq:ham_mu} with the above chemical
potential Eq. \eqref{eq:inverse_mu}, along with the original density
profiles obtained for inhomogeneous hoppings and without chemical
potential. The coincidence between them both is extremely remarkable,
given that Eq. \eqref{eq:inverse_mu} only ensures the similarity
between the highest energy filled mode in both cases.

\begin{figure}
    \centering
    \includegraphics[width=8cm]{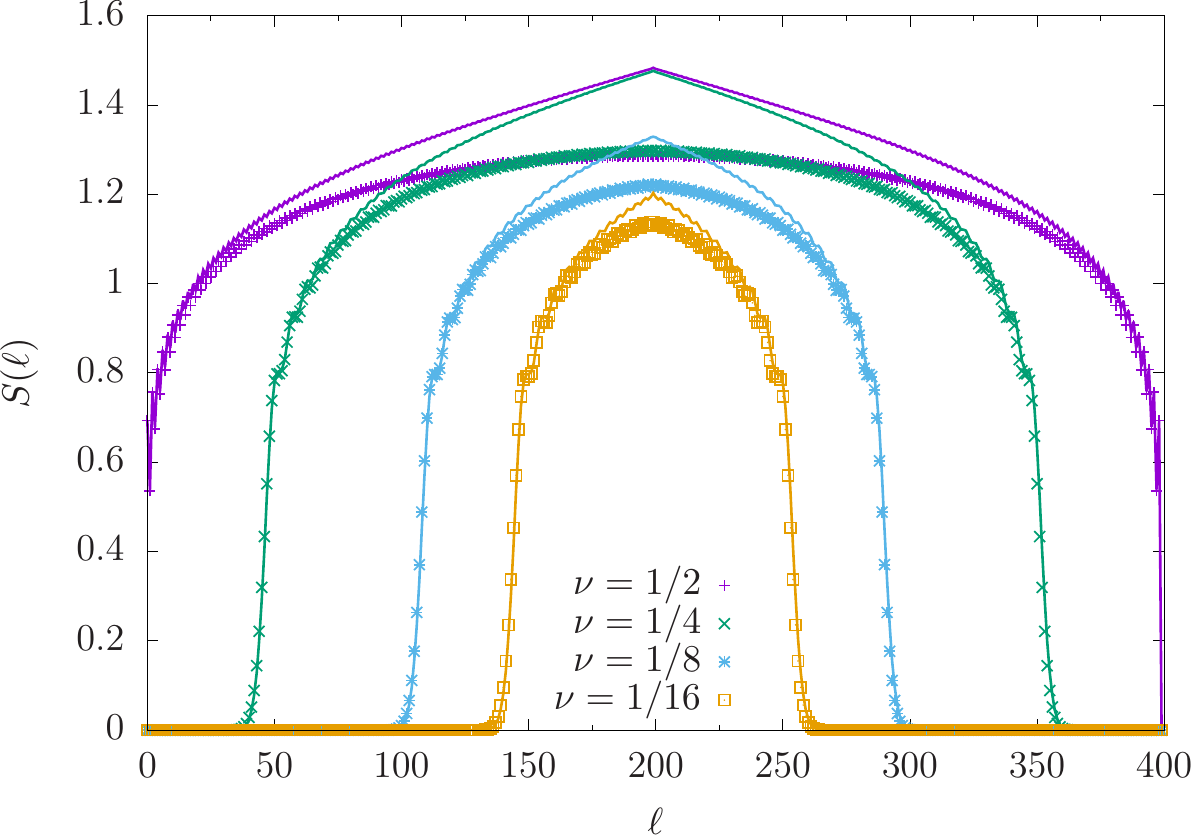}
    \caption{Entanglement entropy $S(\ell)$ of blocks of the form
      $[1,\cdots,\ell]$ of the mimicking GS (points) and the original
      GS (lines), with the same filling fractions for rainbow chains
      with $h=4$ and $N=400$.}
    \label{fig:ee}
\end{figure}

Thus, we are led to ask whether the two GS are the same or
not. Fig. \ref{fig:ee} provides a negative answer to that question. In
it we have shown the entanglement entropies (EE) of blocks
$A=[1,\cdots,\ell]$ as a function of $\ell$ for both states in the
rainbow case, defined as $S(\ell)=-\Tr_A (\rho_A \log\rho_A)$ with
$\rho_A=\Tr_{\bar A} \ket|\Psi>\bra<\Psi|$. As it was shown in
\cite{Finkel.21}, the EE of our states is approximately equal to the
EE of blocks of a shorter system bounded by the turning points at
half-filling \cite{Laguna.17,Samos.21}. Yet, the EE of the GS of the
mimicking system are different, presenting strong similarities to the
EE of homogeneous chains \cite{Calabrese.04}.

%%%%%%%%%%%%%%%%%%%

\section{Conclusions and further work}
\label{sec:conclusions}

A free-fermionic chain without chemical potential and with
inhomogeneous hoppings at half-filling will present an exactly
homogeneous density profile. In the continuum limit, this system
represents a Dirac fermion on a static curved background, with the
lapse function of the metric given by the hopping amplitudes. Yet, if
we move away from half-filling we will notice that the particles
concentrate at the regions with higher hopping amplitudes, and leave
the regions with lower hopping amplitudes empty, a phenomenon that we
have called {\em depletion}.

In the strong inhomogeneity regime, the Dasgupta-Ma renormalization
scheme allows to prove that this should be the case, since the
particles will establish bonds on top of the larger hoppings, either
original or renormalized. Thus, the depletion is exact.

In the weak inhomogeneity regime, we have shown that the associated
single-particle problem is equivalent to a Schrödinger equation on a
different static curved manifold, with the lapse function given by the
{\em square root} of the hopping amplitudes, and a potential
determined by the hopping amplitudes and the filling fraction. Notice
that this shows that the effective system does not show conformal
invariance. Naturally, the laplacian operator must be substituted
with the Laplace-Beltrami operator corresponding to the associated
metric, and the depleted regions correspond to the classically
forbidden areas of this Schrödinger equation. The wavefunctions and
the density profiles can be accurately obtained using a semiclassical
approximation.

It is interesting to ask how this model breaks the conformal symmetry
which is known to hold at half-filling. Indeed, a second order
expansion of the fields is required to find a continuum approximation
to our lattice model, instead of the first-order expansion at
half-filling. The second-order derivative term, which maps into a
laplacian, breaks explicitly the conformal invariance introducing a
length scale, which is inversely proportional to the effective mass.

We may introduce a {\em compensating potential} in our system, through
a chemical potential proportional to the hopping amplitudes, which
exactly cancels the depletion effect and provides exactly homogeneous
density profiles. In this case, the continuous approximation allows us
to conjecture that the system recovers its full conformal invariance.

We have also introduced a {\em mimicking potential}, which provides
exactly the same density profiles away from half filling on a
fermionic chain with homogeneous hoppings. Interestingly, this
mimicking potential is inversely proportional to the hopping
amplitudes. Yet, the associated ground state is not the same as in the
original case, as we have been able to show checking the entanglement
entropies of lateral blocks. The ground states of the compensating and
mimicking systems present interesting challenges which should be
considered in further work.

%%%%%%%%%%%%%%%%%%%%%%%%%%%%

\begin{acknowledgments}
We would like to acknowledge very useful discussions with E. Tonni, A. González-López, F. Finkel , and J.E. Alvarellos. 
This work was funded by the Spanish government through Grants No. PGC2018-095862-B-C21, No. PGC2018-094763-B-I00, No. PID2019-105182GB-I00, No. PID2021-123969NB-I00, No. PID2021-127726NB-I00, the Comunidad de Madrid Grant No. S2018/TCS-4342, the Centro de Excelencia Severo Ochoa Program SEV-2016-0597, the CSIC Research Platform on Quantum Technologies PTI-001 and Comunidad de Madrid under the Multiannual Agreements with UC3M in the line of Excellence of University Professors (Grants No. EPUC3M14 and No. EPUC3M23) in the context of the V Plan Regional de Investigación Científica e Innovación Tecnológica (PRICIT). B.M. acknowledges financial support through Contract No. 2022/167 under the EPUC3M23 line. N.S. acknowledges the financial support from Grant No. QTP2021-03-009 QTEP from the Recovery, Transformation and Resilience Plan (RTRP) national program.
\end{acknowledgments}

%%%%%%%%%%%%%%%%%%%%%%%%%%%%

\end{document}